\documentclass[11pt]{article}

\usepackage{amssymb,graphicx,fullpage,epsfig}

\begin{document}

\title{Symmetry breaking in a model of antigenic variation with immune delay}

\author{K.B. Blyuss\thanks{Corresponding author. Email: k.blyuss@sussex.ac.uk}\hspace{0.2cm} and Y.N. Kyrychko
\\\\Department of Mathematics, University of Sussex,\\Brighton, BN1 9QH, United Kingdom}

\maketitle

\begin{abstract}
Effects of immune delay on symmetric dynamics are investigated within a model of antigenic variation in malaria. Using isotypic decomposition of the phase space, stability problem is reduced to the analysis of a cubic transcendental equation for the eigenvalues. This allows one to identify periodic solutions with different symmetries arising at a Hopf bifurcation. In the case of small immune delay, the boundary of the Hopf bifurcation is found in a closed form in terms of system parameters. For arbitrary values of the time delay, general expressions for the critical time delay are found, which indicate bifurcation to an odd or even periodic solution. Numerical simulations of the full system are performed to illustrate different types of dynamical behaviour.
The results of this analysis are quite generic and can be used to study within-host dynamics of many infectious diseases.
\end{abstract}

\section{Introduction}

Among various strategies employed by pathogens to evade the host immune system, a prominent place is occupied by antigenic variation. Notable examples of pathogens relying on this strategy of immune escape include African Trypanosoma, {\it Plasmodium falciparum}, HIV, several members of {\it Neisseria} family, {\it Haemophilus influenzae} etc. \cite{CS}. Despite the fact that some details of this mechanism are still unclear, the main features of this method of immune evasion are quite universal. The immune system of the host detects potential infection with a pathogen by identifying specific chemical determinants, such as proteins and carbohydrates, known as epitopes, on the surfaces of infected cells. This triggers the differentiation of precursor cells into effector cells, which are then able to eliminate the infection. Some pathogens have evolved to have a wide variety of surface markers (antigens), and by changing the antigens the present on the cell surface, these pathogens can for a long period of time remain unrecognized by the immune system, giving them an opportunity to be transmitted to other hosts. This process of sequentially presenting different antigens in order to avoid the host immune system is known as antigenic variation.

There are several particular ways of implementing antigenic variation. In the case of {\it Trypanosoma brucei}, the organism that causes sleeping sickness, parasite covers itself with a dense homogeneous coat of variant surface glycoprotein (VSG). Genome of {\it T. brucei} has over 1000 genes that control the expression of VSG protein, and switching between them provides the mechanism of antigenic variation \cite{LMRB}. What makes {\it T. brucei}  unique is the fact that unlike other pathogens, whose antigenic variation is typically mediated by DNA rearrangements or transcriptional regulation, activation of VSGs requires recombination of VSG genes into an expression site (ES), which consists of a single {\it vsg} gene flanked by an upstream array of 70 base pair repeats and expression site associated genes (ESAGs). {\it T. brucei} expresses one VSG at any given time, and the active VSG can either be selected by activation of a previously silent ES (and there are up to 20 ES sites), or by recombination of a VSG sequence into the active ES. The precise mechanism of VSG switching has not been completely identified yet, but it has been suggested that the ordered appearance of different VSG variants is controlled by differential activation rates and density-dependent parasite differentiation \cite{LMRB,SSBM}.

For the malaria agent {\it P. falciparum}, the main target of immune response is {\it Plasmodium falciparum} erythrocyte membrane protein-1 (PfEMP1), which is expressed from a diverse family of {\it var} genes, and each parasite genome contains approximately 60 {\it var} genes encoding different PfEMP1 variants \cite{Ga}. The {\it var} genes are expressed sequentially in a mutually exclusive manner, and this switching between expression of different {\it var} gene leads to the presentation of different  variant surface antigens (VSA) on the surface of infected erythrocyte, thus providing a mechanism of antigenic variation \cite{BBMLR,New}. In all cases of antigenic variation, host immune system has to go through a large repertoire of antigenic variants, and this provides parasites with enough time to get transmitted to another host or cause a subsequent infection with a different antigenic variant in the same host. 

Despite individual differences in the molecular implementation of antigenic variation, such as, gene conversion, site-specific DNA inversions, hypermutation etc., there are several features common to the dynamics of antigenic variation in all pathogens \cite{CS}. These include ordered and often sequential appearance of parasitemia peaks corresponding to different antigenic variants, as well as certain degree of cross-reactivity. Several mathematical models have been put forward that aim to explain various aspects of antigenic variation. Agur {\it et al.} \cite{Ag} have studied a model of antigenic variation of African trypanosomes which suggests that sequential appearance of different antigenic variants can be explained by fitness differences between single- and double-expressors - antigenic variants that express one or two VSGs. However, this idea is not supported by the experimental evidence arising from normal {\it in vivo} growth and reduced immunogenicity of artificially created double expressors \cite{MJ}. Frank \cite{FB} has suggested a model that highlights the importance of cross-reactivity between antigenic variants in facilitating optimal switching pattern that provides sequential dominance and extended infection. Antia {\it et al.} \cite{ANA} have considered variant-transcending immunity as a basis for competition between variants, which can promote oscillatory behaviour, but this failed to induce sequential expression. Many other mathematical models of antigenic variation have been proposed and studied in the literature, but the discussion of their individuals merits and limitations is beyond the scope of this work.

The model considered in this paper is a modification of the model proposed by Recker {\it et al.} \cite{Re04} (to be referred to as Recker model), which postulates that in addition to a highly variant-specific immune response, the dynamics of each variant is also affected by cross-reactive immune responses against a set of epitopes not unique to this variant. This assumption implies that each antigenic variant experiences two types of immune responses: a long-lasting immune response against epitopes unique to it, and a transient immune response against epitopes that it shares with other variants. The main impact of this model lies in its ability to explain a sequential appearance of antigenic variants purely on the basis of cross-reactive inhibitory immune responses between variants sharing some of their epitopes, without the need to resort to variable switch rates or growth rates (see Gupta \cite{Gu} for a discussion of several clinical studies in Ghana, Kenya and India, which support this theory). 

In the case of non-decaying long-lasting immune response, numerical simulations in the original paper \cite{Re04} showed that eventually all antigenic variants will be cleared by the immune system, with specific immune responses reaching protective levels preventing each of the variants from showing up again. Blyuss and Gupta \cite{BG} have demonstrated that the sequential appearance of parasitemia peaks during such immune clearance can be explained by the existence of a hypersurface of equilibria in the phase space of the system, with individual trajectories approaching this hypersurface and then being pushed away along stable/unstable manifolds of the saddle-centres lying on the hypersurface. They also numerically analysed robustness of synchronization between individual variants. Under assumption of perfect synchrony, when all variants are identical to each other, Recker and Gupta \cite{RG} have analysed peak dynamics and threshold of chronicity, while Mitchell and Carr \cite{MC1} have considered the case of slowly decaying specific immune response. De Leenheer and Pilyugin \cite{LP} have replaced linear growth of antigenic variants in the original model by the logistic growth, and have studied the effects of various types of cross-reactivity on the dynamics, ranging from no cross-reactivity to partial and complete cross-immunty.

It is known that time delay in the immune response can have a profound effect on the dynamics of host-parasite interactions and the host ability to eliminate infection. Several models have studied mathematically the effects of time delay on the immune dynamics and possible onset of oscillatory behaviour \cite{BMV,Mar,MZH,MB}. For the Recker model, Mitchell and Carr have investigated the effect of time delay in the development of immune response in the case of complete synchrony between antigenic variants \cite{MC1}, and they have also investigated the appearance of synchronous and asynchronous oscillations \cite{MC2} in the case of global coupling between variants (referred to as "perfect cross immunity" in \cite{LP}).

In this paper, we use methods of equivariant bifurcation theory for delay differential equations to study the dynamics of a fully symmetric state in the Recker model. Stability analysis of the appropriate characteristic equation will show that under certain conditions on parameters, this state can undergo Hopf bifurcation, giving rise to different types of stable periodic solutions. The outline of this paper is as follows. In the next section we introduce the mathematical model of antigenic variation with time delay in the immune response and discuss its main properties. Section 3 discussed different steady states and derives the transcendental characteristic equation, which determines the stability of the fully symmetric state. In Sections 4 and 5 we analyse the case of small and arbitrary time delay, respectively, and find the boundary of Hopf bifurcation in terms of system parameters and the immune delay. Section 6 contains numerical simulations of the model and analysis of the symmetry properties of different periodic solutions. The paper concludes in Section 7 with the discussion of results and an outlook.

\section{Mathematical model}

In this section, the model of the immune response to malaria is presented together with some facts about the dynamics of this system. Following Recker {\it et al.} \cite{Re04}, we assume that each antigenic variant $i$ consists of a single unique major epitope, that elicits a long-lived (specific) immune response, and also of several minor epitopes that are not unique to the variant. Assuming that all variants have the same net
growth rate $\phi$, their temporal dynamics is described by the
equation
\begin{equation}\label{yeq}
\frac{dy_i}{dt}=y_i(\phi -\alpha z_i-\alpha' w_i),
\end{equation}
where $\alpha$ and $\alpha'$ denote the rates of variant
destruction by the long-lasting immune response $z_i$ and by the
transient immune response $w_i$, respectively, and index $i$ spans
all possible variants. 

It is known that the discovery of infection by immune receptors does not instantaneously lead to the development of the corresponding immune response \cite{Mar}. To include this feature explicitly in the model, we introduce time delay $\tau$ as the time delay that elapses between changes in parasitemia and production of the corresponding immune effectors \cite{MC1,MC2}. For simplicity, it will be assumed that this time delay is the same for both specific and cross-reactive immune responses. The dynamics of the variant-specific immune response can be written in its simplest form as
\begin{equation}\label{zeq}
\frac{dz_i}{dt}=\beta y_{i}(t-\tau)-\mu z_{i}(t),
\end{equation}
where $\beta$ is the proliferation rate, $\mu$ is the
decay rate of the immune response, and $\tau$ is the above-mentioned time delay in the immune response. Finally, the transient (cross-reactive) immune response can be described by a minor modification of the above equation (\ref{zeq}):
\begin{equation}\label{weq}
\frac{dw_i}{dt}=\beta'\sum_{j\sim i} y_j(t-\tau)-\mu' w_i,
\end{equation}
where the sum is taken over all variants sharing the epitopes with
the variant $y_i$. We shall use the terms long-lasting and
specific immune response interchangeably, likewise for transient
and cross-reactive.

\begin{figure}
\hspace{4cm}
\epsfig{file=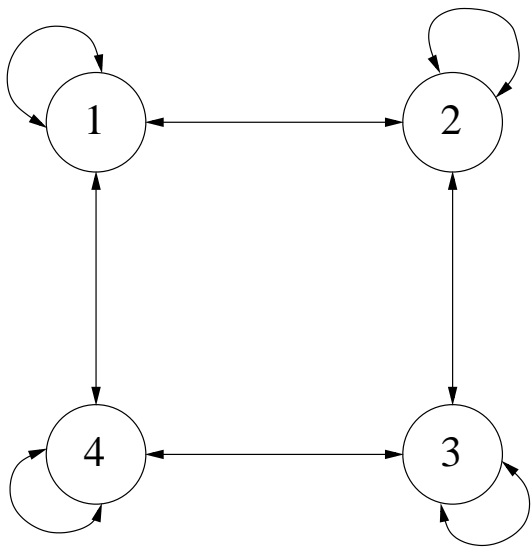,width=7cm}
\caption{Interaction of malaria variants in the case of two minor epitopes with two variants in
each epitope.}\label{Vard}
\end{figure}

The above system can be formalized with the help of an
adjacency matrix $T$, whose entries $T_{ij}$ are equal to one if
the variants $i$ and $j$ share some of their minor epitopes and
equal to zero otherwise. Obviously, the matrix $T$ is always a
symmetric matrix. Prior to constructing this matrix it is
important to introduce a certain ordering of the variants
according to their epitopes. To illustrate this, suppose we have a system of two minor epitopes with two
variants in the each epitope. In this case, the total number of variants is four, and they are enumerated as follows
\begin{equation}\label{var4}
\begin{array}{l}
1\hspace{1cm}11\\
2\hspace{1cm}12\\
3\hspace{1cm}22\\
4\hspace{1cm}21
\end{array}
\end{equation}
The diagramme of interactions between these antigenic variants is shown in Fig.~\ref{Vard}. It is clear that for a system
of $m$ minor epitopes with $n_{i}$ variants in each epitope, the total number of variants is given by
\begin{equation}
N=\prod_{i=1}^{m}n_{i}.
\end{equation}

After the ordering of variants has been fixed, it is straightforward
to construct the connectivity matrix $T$ of variant
interactions. For the particular system of variants (\ref{var4}),
this matrix has the form
\begin{equation}\label{Amat}
T=\left(
\begin{array}{cccccc}
1&1&0&1\\
1&1&1&0\\
0&1&1&1\\
1&0&1&1
\end{array}
\right).
\end{equation}
For the rest of the paper we will concentrate on the
case of two minor epitopes, but the results can be
generalized to larger systems of antigenic variants. 
Using the connectivity matrix one can rewrite the system
(\ref{yeq})-(\ref{weq}) in a vector form
\begin{equation}\label{vs}
\frac{d}{dt} \left(\begin{array}{l}{\bf y}\\
{\bf z}\\{\bf w}\end{array}\right) =F({\bf y},{\bf z},{\bf w})=
\left\{
\begin{array}{l}
{\bf y}(\phi{\bf 1}_{N}-\alpha{\bf z}-\alpha'{\bf w}),\\
\beta{\bf y}(t-\tau)-\mu{\bf z},\\
\beta'T{\bf y}(t-\tau)-\mu'{\bf w},
\end{array}
\right.
\end{equation}
where ${\bf y}=(y_1,y_2,...,y_{N})$ etc., ${\bf 1}_{N}$ denotes a
vector of the length $N$ with all components equal to one, and in the right-hand side of the first equation multiplication is taken
to be entry-wise so that the output is a vector again. The above system has to be augmented by appropriate initial conditions,
which are taken to be
\[
\begin{array}{l}
{\bf y}(\theta)={\bf \psi}(\theta)\geq 0,\theta\in[-\tau,0],\\
{\bf z}(0)\geq 0,{\bf w}\geq 0.
\end{array}
\]
with the history function $\psi(\theta)\in C([-\tau,0],\mathbb{R}^N)$, where $C([-\tau,0],\mathbb{R}^N)$ denotes the Banach space of continuous mappings from $[-\tau,0]$ into $\mathbb{R}^{N}$ equipped with the supremum norm $\|\psi\|=\sup_{-\tau\leq\theta\leq 0}|\psi(\theta)|$ for $\psi\in C([-\tau,0],\mathbb{R}^N)$, where $|\cdot |$ is the usual Euclidean norm on $\mathbb{R}^N$. Using the same argument as in \cite{BG}, it is possible to show that with these initial conditions, the system (\ref{vs}) is well-posed, i.e. its solutions remain non-negative for all $t\geq 0$.

We will assume that cross-reactive immune responses develop at a slower rate than specific immune responses, have a shorter life time, and are less efficient in destroying the infection. This implies the following relations between the system parameters
\begin{equation}\label{Pars}
\alpha'\leq \alpha,\hspace{0.5cm}\mu\leq \mu',\hspace{0.5cm}\beta'\leq \beta.
\end{equation}

In terms of symmetry in the network of interactions between different antigenic variants, in the case of two minor epitopes with $m$ variants in the first epitope and $n$ variants in the second, the system (\ref{vs}) is equivariant with respect to the following symmetry group \cite{B11,BG}
\begin{equation}
\Gamma=\left\{
\begin{array}{l}
{\bf S}_{m}\times {\bf S}_{n},\mbox{ }m\neq n,\\
{\bf S}_{m}\times {\bf S}_{m}\times{\bf Z}_{2},\mbox{
}m=n.
\end{array}
\right.
\end{equation}
Here, ${\bf S}_m$ denotes the symmetric group of all permutations in a network of $n$ nodes with an all-to-all coupling, and ${\bf Z}_2$ is the cyclic group of order 2, which corresponds to rotations by $\pi$.

The above construction can be generalized in a straightforward way to a larger number of minor epitopes. System (\ref{vs}) provides an interesting example of a linear coupling, which for $N>4$ does not reduce to known symmetric configurations, such as diffusive, star or all-to-all coupling \cite{Pe}. A really important aspect is that two systems of antigenic variants with the same total number of variants $N$ may have different symmetry properties as described by the symmetry group $\Gamma$ depending on $m$ and $n$, such that $N=mn$.

\section{Steady states}

In the particular case of non-decaying specific immune response $(\mu=0)$ and instantaneous immune response ($\tau=0$), steady states of the system (\ref{vs}) are not isolated but rather form an $N$-dimensional hypersurface $H_{0}=\{({\bf y},{\bf z},{\bf w})\in\mathbb{R}^{N}:{\bf y}={\bf w=0}_N\}$ in the phase space \cite{BG}. Linearization near each equilibrium on this hypersurface has the eigenvalues $(-\mu')$ of multiplicity $N$, zero of multiplicity $N$, and the the rest of the spectrum is given by
\[
\phi-\alpha z_1,\phi-\alpha z_2,\ldots,\phi-\alpha z_N.
\]
This suggests that the hypersurface consists of saddles and stable nodes, and besides the original symmetry of the system it possesses an additional translational symmetry along the ${\bf z}$ axes. Furthermore, stability of equilibria on the hypersurface does not depend on the time delay $\tau$. The existence of this hypersurface of equilibria in the phase space leads to a particular behaviour of phase trajectories, which mimics the occurrence of sequential parasitimea peaks in the immune dynamics of malaria \cite{BG,Re04}.

When $\mu>0$, the structure of the phase space of the system (\ref{vs}) and its steady states is drastically different. Now, the only symmetry present is the original symmetry $\Gamma$, and the hypersurface of equilibria $H_{0}$ disintegrates into just two distinct points: the origin $\mathcal{O}$, which is always a saddle, and the fully symmetric equilibrium
\begin{equation}
\begin{array}{l}
E=(Y{\bf 1}_{N},Z{\bf 1}_{N},W{\bf 1}_{N}), \hspace{0.5cm}\mbox{where}\\\\
\displaystyle{
Y=\frac{\phi\mu\mu'}{\alpha\beta\mu'+\alpha'n_{c}\beta'\mu},\hspace{0.5cm}
Z=\frac{\phi\beta\mu'}{\alpha\beta\mu'+\alpha'n_{c}\beta'\mu},\hspace{0.5cm}
W=\frac{\phi\mu n_{c}\beta'}{\alpha\beta\mu'+\alpha'n_{c}\beta'\mu}}.
\end{array}
\end{equation}
Here $n_{c}$ is the total number of connections for each variant, which in the case of two minor epitopes with $m$ variants in the first epitope and $n$ variants in the second, is equal to $n_{c}=m+n-1$. It has been previously shown that in the absence of time delay, the fully symmetric equilibrium $E$ can undergo Hopf bifurcation as the system parameters are varied \cite{B11,BG}. It is worth noting that if one assumes all variants to be exactly the same, the system collapses onto a system with just 3 dimensions, but in this case it is possible to show that the fully symmetric equilibrium is always stable for $\tau=0$ \cite{RG} and can have a Hopf bifurcation for $\tau>0$ \cite{MC1}. Besides the origin and the fully symmetric steady state, system (\ref{vs}) has $2^N-1$ other steady states, all of which are unstable \cite{B11,BG}.

In order to understand the structure of the solution that arises from the Hopf bifurcation of the fully symmetric steady state $E$,  we concentrate on a specific connectivity matrix $T$ (\ref{var4}) corresponding to a particular example of a system of two epitopes with two variants in each epitope, as shown in Fig.~\ref{Vard}. In this case, the system (\ref{vs}) is equivariant under the action of a dihedral group ${\bf D}_{4}$, which is a symmetry group of a square \cite{B11}. We can use the subspaces associated with four one-dimensional irreducible representations of this group to perform an isotypic decomposition of the full phase space $\mathbb{R}^{12}$ as follows \cite{B11,DM,Swift}:
\begin{equation}\label{IsD}
\mathbb{R}^{12}=\{(1,1,1,1),(1,-1,1,-1),(1,0,-1,0),(0,1,0,-1)\}^3.
\end{equation}
The characteristic matrix of the linearization of system (\ref{vs}) near a fully symmetric steady state $E$ has the block form
\[
J(\lambda,\tau)=\left(
\begin{array}{ccc}
-\lambda{\bf 1}_4 &-\alpha {\bf 1}_4 &-\alpha' {\bf 1}_4\\
\beta e^{-\lambda\tau} {\bf 1}_4 & -(\lambda+\mu) {\bf 1}_4& {\bf 0}_4\\
\beta' e^{-\lambda\tau}  T & {\bf 0}_4 & -(\lambda+\mu') {\bf 1}_4
\end{array}
\right),
\]
where ${\bf 0}_4$ and ${\bf 1}_4$ are $4\times 4$ zero and unit matrices, and $T$ is the connectivity matrix (\ref{Amat}). Rather than compute stability eigenvalues directly from this $12\times 12$ matrix, we use isotypic decomposition (\ref{IsD}) to rewrite this characteristic matrix in the block-diagonal form \cite{B11,Swift}
\begin{equation}
\Delta(\lambda,\tau)=\left(
\begin{array}{cccc}
M+2N & {\bf 0}_3&{\bf 0}_3&{\bf 0}_3\\
{\bf 0}_3&M-2N&{\bf 0}_3&{\bf 0}_3\\
{\bf 0}_3&{\bf 0}_3&M&{\bf 0}_3\\
{\bf 0}_3&{\bf 0}_3&{\bf 0}_3&M
\end{array}
\right),
\end{equation}
where
\begin{equation}
M=\left(
\begin{array}{ccc}
-\lambda&-\alpha&-\alpha'\\
\beta e^{-\lambda\tau}&-(\lambda+\mu)&0\\
\beta' e^{-\lambda\tau}&0&-(\lambda+\mu')
\end{array}
\right),\hspace{0.5cm}
N=\left(
\begin{array}{ccc}
0&0&0\\
0&0&0\\
\beta' e^{-\lambda\tau}&0&0
\end{array}
\right).
\end{equation}
Here, matrix $M$ is associated with self-coupling, and $N$ is associated with nearest-neighbour coupling. From the perspective of stability analysis, eigenvalues of the characteristic matrix $\Delta(\lambda,\tau)$ are determined as the the roots of the corresponding characteristic equation
\begin{equation}\label{Deq}
\det \Delta(\lambda,\tau)=\left(\det[\Delta_1(\lambda,\tau)]\right)^2\cdot \det[\Delta_2(\lambda,\tau)]\cdot\det[\Delta_3(\lambda,\tau)]=0,
\end{equation}
where
\[
\Delta_{1,2,3}(\lambda,\tau)=\left(
\begin{array}{ccc}
-\lambda&-\alpha&-\alpha'\\
\beta e^{-\lambda\tau}&-(\lambda+\mu)&0\\
B'_{1,2,3}  e^{-\lambda\tau}&0&-(\lambda+\mu')
\end{array}
\right),
\]
with  $B'_{1,2,3}=\beta',3\beta',-\beta'$ for matrices $M$, $M+2N$ and $M-2N$, respectively. Stability is now determined by the roots of the corresponding characteristic equation
\begin{equation}\label{char}
\lambda^3+(\mu+\mu')\lambda^2+\mu\mu'\lambda+\lambda(\alpha\beta+\alpha' B')e^{-\lambda\tau}+(\alpha\beta\mu'+\alpha' B'\mu)e^{-\lambda\tau}=0.
\end{equation}
This equation can be rewritten in the form
\begin{equation}\label{CE}
\lambda^3+A\lambda^2+B\lambda+C\lambda e^{-\lambda\tau}+De^{-\lambda\tau}=0,
\end{equation}
where
\[
A=\mu+\mu',\hspace{0.3cm}B=\mu\mu',\hspace{0.3cm}C=\alpha\beta+\alpha' B',\hspace{0.3cm}D=\alpha\beta\mu'+\alpha' B'\mu.
\]
When the immune response is instantaneous ($\tau=0$), the above equation simplifies to
\[
\lambda^3+A\lambda^2+(B+C)\lambda+D=0,
\]
and one can use the Routh-Hurwitz criterion to deduce that in the cases $B'=\beta'$ and $B'=3\beta'$, the fully symmetric steady state $E$ never loses stability, as the eigenvalues remain in the left complex half plane for all values of the parameters. When $B'=-\beta'$, the steady state $E$ can undergo
Hopf bifurcation at
\begin{equation}\label{HopfB}
\alpha'_{H}=\frac{\alpha\beta\mu+\mu\mu'(\mu+\mu')}{\beta'\mu'},
\end{equation}
thus giving rise to an odd periodic orbit, where variants 1 and 3 are synchronized and half a period out-of-phase with variants 2 and 4, i.e. each variant is $\pi$ out of phase with its nearest neighbours \cite{B11}. Another possibility is for the steady state $E$ to undergo a steady state bifurcation at
\[
\alpha'_{SS}=\frac{\alpha\beta\mu'}{\beta'\mu},
\]
provided $\mu'<\alpha\beta\mu/\left(\alpha\beta-\mu^2\right)$, but due to restrictions on parameters (\ref{Pars}), this cannot happen.

Cubic quasi-polynomial equations similar to (\ref{CE}) have been previously studied in several applied contexts, such as, models of business cycles \cite{Cai}, testosterone secretion \cite{RW1,RW2}, or neural networks with bidirectional associative memory \cite{SHW}. In each of those cases, analyses of the appropriate characteristic equation allowed one to find restrictions on system parameters and the time delay, which lead to the occurrence of Hopf bifurcation.

We will use the results of equivariant bifurcation theory for delay differential equations to analyse symmetry properties of possible solutions arising at the Hopf bifurcation. While the effects of symmetry on the dynamics of Hopf bifurcation in systems without time delay have been known for quite a long time \cite{AKS,GS2,GSS,Swift}, it is only in the last 10-15 years that these results have been adapted to the analysis of delay and functional differential equations. In a series of papers, Wu and co-workers extended the theory of equivariant Hopf bifurcation to systems with time delays and employed equivariant degree theory to study existence, multiplicity and global continuation of symmetric periodic solutions \cite{KVW1,KVW2,KW1,Wu98}. The results of this analysis have been subsequently applied to the studies of Hopf bifurcation in a number of symmetric models of coupled oscillators with delayed coupling \cite{BC,CYB,FW,GH,YC04}.

The strategy now is to consider when the eigenvalues of the characteristic equation (\ref{CE}) cross the imaginary axis with non-zero speed, giving rise to a Hopf bifurcation. One has to separately consider the cases $B'=3\beta'$, $B'=-\beta'$, and $B'=\beta'$, as these correspond to a Hopf bifurcation in the even, odd and $V_4$ subspaces, respectively \cite{Swift}. In the case of bifurcation in the even subspace, the periodic solution that appearing will be an even periodic orbit, which has the full original symmetry ${\bf D}_4$ and is characterized by all variants oscillating in perfect synchrony with each other. An odd solution, also called anti-phase solution, corresponds to a bifurcation in the odd subspace, and has variants 1 and 3 oscillating synchronously and half of a periodic out-of-phase with variants 2 and 4. Finally, when the bifurcation takes place in the $V_4$ subspace, this is known as the Hopf bifurcation with symmetry \cite{Swift}; as is clear from (\ref{Deq}), in this case, two pairs of complex conjugate eigenvalues simultaneously cross the imaginary axis, and this can give rise to a number of different periodic behaviours, including edge and vertex oscillations, as well as discrete travelling waves \cite{Swift}. By finding the minimum value of time delay, at which Hopf bifurcation occurs for one of the three possible values of $B'$, one can identify the type of periodic solution that will appear at the corresponding Hopf bifurcation.

\section{Small immune delay}

We begin our analysis of stability by considering the case when the time delay in the development of immune response is small $0<\tau\ll 1$. In this case, one can write $e^{-\lambda\tau}\approx 1-\lambda\tau$, which transforms the characteristic equation (\ref{CE}) into a regular cubic equation
\begin{equation}\label{cubic}
\lambda^{3}+a_1(\tau)\lambda^2+a_2(\tau)\lambda+a_3=0,
\end{equation}
where
\begin{equation}\label{ccoef}
\begin{array}{l}
a_1(\tau)=\mu+\mu'-\tau(\alpha\beta+\alpha' B'),\\\\
a_2(\tau)=\mu\mu'+\alpha\beta+\alpha' B'-\tau(\alpha\beta\mu'+\alpha' B'\mu),\\\\
a_3=\alpha\beta\mu'+\alpha' B'\mu.
\end{array}
\end{equation}
While finding the roots of the cubic equation (\ref{cubic}) is still too cumbersome, one can resort to the Routh-Hurwitz criterion to establish the conditions for stability of fully symmetric steady state $E$. These conditions are as follows
\[
\begin{array}{l}
a_i>0,\hspace{0.3cm}i=1,2,3,\\
a_1 a_2-a_3>0.
\end{array}
\]
For $\tau$ small but different from zero, we can find that the Hopf bifurcation will occur when 
\begin{equation}\label{Hsmall}
\begin{array}{l}
a_1(\tau)>0,\hspace{0.5cm}a_2(\tau)>0,\hspace{0.5cm}a_3>0,\mbox{ and}\\
a_1(\tau)a_2(\tau)=a_3,
\end{array}
\end{equation}
which gives the critical value of the time delay as
\begin{equation}\label{tauh}
\tau_H=\frac{P-\sqrt{P^2-4QR}}{2Q},
\end{equation}
with
\[
\begin{array}{l}
Q=(\alpha\beta\mu'+\alpha' B'\mu)(\alpha\beta+\alpha' B'),\\\\
P=(\alpha\beta+\alpha' B')(\alpha\beta+\alpha' B'+\mu\mu')+(\mu+\mu')(\alpha\beta\mu'+\alpha' B'\mu),\\\\
R=\alpha\beta\mu+\alpha' B'\mu'+\mu\mu'(\mu+\mu').
\end{array}
\]
If we define a characteristic polynomial as
\[
g(\lambda,\tau)=\lambda^{3}+a_1(\tau)\lambda^2+a_2(\tau)\lambda+a_3,
\]
the characteristic equation (\ref{cubic}) at $\tau=\tau_H$ becomes
\[
g(\lambda,\tau_H)=\lambda^{3}+a_1(\tau_H)\lambda^2+a_2(\tau_H)\lambda+a_1(\tau_H)a_2(\tau_H)=0.
\]
The eigenvalues of (\ref{cubic}) at $\tau=\tau_H$ can be readily found as
\[
\lambda_1(\tau_H)=-a_1(\tau_H)<0,
\]
and
\[
\lambda_{2,3}(\tau_H)=\pm i\omega,\hspace{0.5cm}\omega=\sqrt{a_2(\tau_H)},
\]
where $\omega$ is the Hopf frequency. To establish the occurrence of a Hopf bifurcation, we need to show that $\displaystyle{{\rm Re}(d\lambda/d\tau)|_{\tau=\tau_H}>0}$. Since $dg/d\tau=(\partial g/\partial\tau)+(\partial g/\partial\lambda)(d\lambda/d\tau)=0$, we have
\[
\frac{d\lambda}{d\tau}=-\frac{\partial g}{\partial\tau}\Big/\frac{\partial g}{\partial\lambda}=-\frac{-(\alpha\beta+\alpha' B')\lambda^2-a_1(\tau)a_2(\tau)\lambda}{3\lambda^2+2a_1(\tau)\lambda+a_2(\tau)}.
\]
Evaluating this at $\tau=\tau_H$ gives
\begin{equation}
\frac{d\lambda(\tau)}{d\tau}\Bigg|_{\tau=\tau_H}\!\!\!\!\!\!\!\!=-\frac{\left[a_2(\tau_H)(\alpha\beta+\alpha' B')-a_1(\tau_H)a_2(\tau_H)^{3/2}i\right][-a_2(\tau_H)-a_1(\tau_H)\sqrt{a_2(\tau_H)}i]}{2a_2(\tau_H)[a_1(\tau_H)+a_2(\tau_H)]}.
\end{equation}

\newpage
\begin{figure}[h]
\epsfig{file=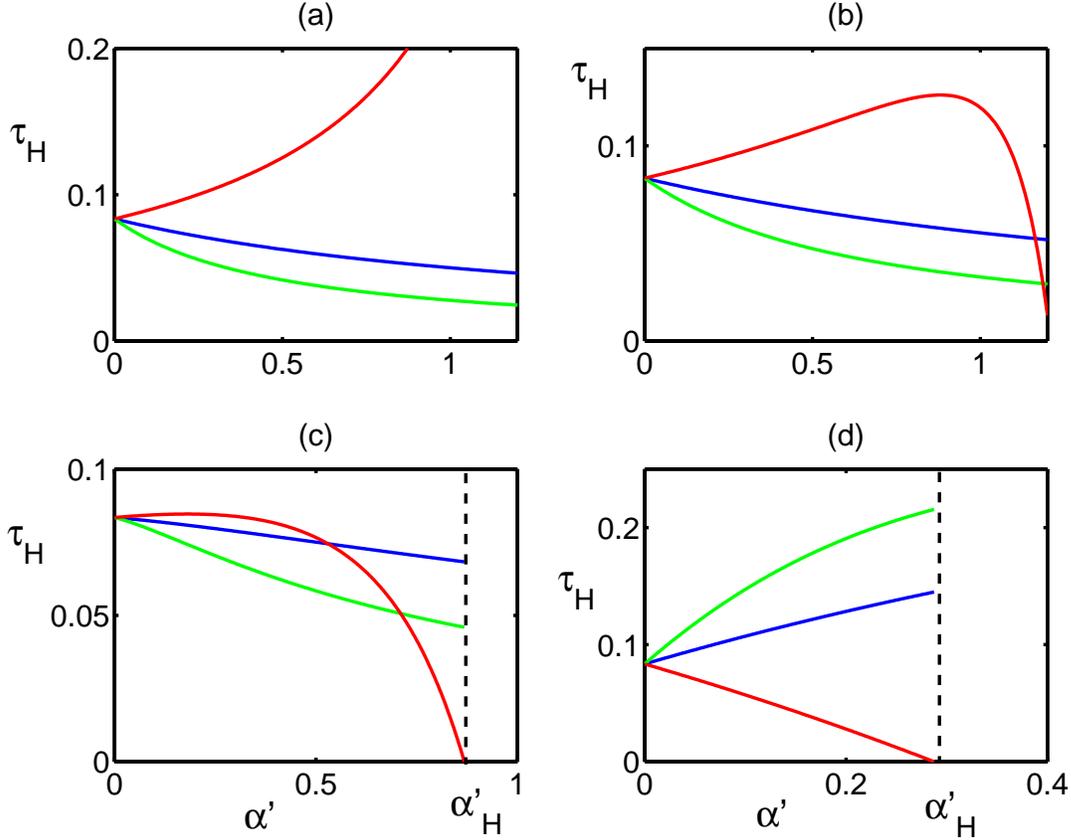,width=16cm}
\caption{Critical time delay $\tau_H$ (\ref{tauh}) at the Hopf bifurcation as a function of $\alpha'$. Parameter values are $\beta=1$, $\alpha=1.2$, $\mu=0.1$, $\beta'=0.8$, (a) $\mu'=0.1$, (b) $\mu'=0.127$, (c) $\mu'=0.18$, (d) $\mu'=1.2$. The colour corresponds to $B'=\beta'$ (blue), $B'=3\beta'$ (green), and $B'=-\beta'$ (red).}\label{SDfig}
\end{figure}

\noindent The real part can be found as
\begin{equation}
\left[\frac{d{\rm Re}(\lambda(\tau))}{d\tau}\right]\Bigg|_{\tau=\tau_H}=\frac{a_2(\tau_H)[a^2_1(\tau_H)+\alpha\beta+\alpha' B']}{2[a_1(\tau_H)+a_2(\tau_H)]}>0.
\end{equation}
When $B'=\beta'$ or $B'=3\beta'$, this is true since all parameters are positive; for $B'=-\beta'$, this inequality holds due to $\alpha'\leq\alpha$ and $\beta'\leq\beta$, as required by (\ref{Pars}). This implies that in all three cases, the eigenvalues $\lambda_{2,3}$ cross the imaginary axis at $\tau=\tau_H$ with a positive velocity, which implies the existence of a Hopf bifurcation at $\tau=\tau_H$. These finding can be summarized in the following theorem.\\

\noindent {\bf Theorem 1.} {\it For sufficiently small values of the time delay $0<\tau\ll 1$, the fully symmetric steady state $E$ is stable provided $\tau<\tau_H$ as defined in (\ref{tauh}) and  unstable for $\tau>\tau_H$. At $\tau=\tau_H$, this steady state undergoes Hopf bifurcation. If the minimal value of $\tau_H$ corresponds to $B'=3\beta'$, the bifurcating solution will be an even periodic orbit; if it corresponds to $B'=-\beta'$, the bifurcating solution will be an odd periodic orbit; if it corresponds to $B'=\beta'$, the bifurcating periodic orbit will lie in the $V_4$ subspace. If $\alpha'_H$ as defined in (\ref{HopfB}) satisfies $\alpha'_H<\alpha$, then for $\alpha'=\alpha'_H$, the steady state $E$ undergoes Hopf bifurcation to an odd periodic orbit for zero time delay $\tau$.}\\

Figure~\ref{SDfig} illustrates how the critical time delay $\tau_H$ at the Hopf bifurcation varies with the rate $\alpha'$ of variant destruction by transient immune response, and also with the decay rate of the transient immune response $\mu'$. The type of bifurcating periodic orbit is determined by which of the matrices $M$, $M\pm N$ will become unstable first as the time delay $\tau$ increases. One can observe that when the death rate of cross-reactive immune response $\mu'$ is sufficiently close to that of the specific immune response $\mu$, i.e. when the cross-reactive immune response is sufficiently long-lasting, the only possible periodic solution the steady state $E$ can bifurcate to is the even solution (see Fig. (a)), in which all antigenic variants are behaving identically to each other. As the lifetime of transient immune response gets shorter (i.e. the rate $\mu'$ increases), there appears a range of $\alpha'$ values shown in Figs. (b) and (c), for which the steady state $E$ can also bifurcate to an odd periodic orbit, in which variants 1 and 3 are synchronized and half-a-period out of phase with variants 2 and 4. As $\mu'$ increases further, the possibility of bifurcating into an even solution completely disappears (see Fig. (d)), and the only remaining possibility is a bifurcation to an odd solution for all admissible values of $\alpha'$. It is worth noting that for higher values of $\mu'$, the range of possible values of $\alpha'$ for which Hopf bifurcation can occur, is bounded by $\alpha'_H$ given in (\ref{HopfB}), for which the steady state $E$ bifurcates to an odd periodic orbit for $\tau=0$.

\section{The case of general delay}

In the situation when the immune delay $\tau$ is not small, the approximation used in the previous section is not valid, and one has to consider the full equation (\ref{CE}). Before proceeding to the analysis of the full characteristic equation, it is worth mentioning a general result of Hale \cite{Hale}, which guarantees {\it absolute stability} of a delay system - this is a case when the real parts of all eigenvalues remain negative {\it for all} values of time delay, i.e. effectively independent of the delay. This is achieved when the corresponding ODE system is asymptotically stable, and the characteristic equation has no purely imaginary roots.

Now we consider the characteristic equation (\ref{CE}) and analyse its roots in order to identify possible parameter regimes when the fully symmetric steady state $E$ can lose its stability. First of all, due to biological restrictions on system parameters, the coefficient $D$ in the equation (\ref{CE}) is always positive, and this implies that $\lambda=0$ is a not a root of the equation. Therefore, the only way how the steady state $E$ can lose its stability is through a Hopf bifurcation at some $\tau=\tau_0$, when a pair of eigenvalues of the equation (\ref{CE}) crosses the imaginary axis. Let us assume that when $\tau=0$, the steady state $E$ is stable, which is always the case for $B'=\beta'$ and $B'=3\beta'$, and is also true for $B'=-\beta'$ provided $\alpha'<\alpha'_H$, where $\alpha'_H$ is given in (\ref{HopfB}). In order to find out when the Hopf bifurcation can occur as the time delay $\tau$ increases from zero, we look for solutions of equation (\ref{CE}) in the form $\lambda=i\omega$ $(\omega>0)$. Such a solution would be a root of (\ref{CE}) if and only if $\omega$ satisfies
\[
-i\omega^3-A\omega^2+iB\omega+(Ci\omega+D)(\cos\omega\tau-i\sin\omega\tau)=0.
\]
Separating the real and imaginary parts, we have
\begin{equation}\label{ReIm}
\left\{
\begin{array}{l}
A\omega^2=D\cos\omega\tau+C\omega\sin\omega\tau,\\
\omega(B-\omega^2)=D\sin\omega\tau-C\omega\cos\omega\tau.
\end{array}
\right.
\end{equation}
Squaring and adding these two equations yields a single polynomial equation for the Hopf frequency $\omega$:
\begin{equation}\label{OmEq}
\omega^6+\left(A^2-2B\right)\omega^4+\left(B^2-C^2\right)\omega^2-D^2=0.
\end{equation}
Let $x=\omega^2$, and also
\begin{equation}\label{Cpars}
c_1=A^2-2B,\hspace{0.3cm}c_2=B^2-C^2,\hspace{0.3cm}c_3=-D^2.
\end{equation}
Now, introducing function $f(x)$ as
\begin{equation}\label{fdef}
f(x)=x^3+c_1x^2+c_2x+c_3,
\end{equation}
we can rewrite equation (\ref{OmEq}) as
\begin{equation}\label{Heq}
f(x)=x^3+c_1x^2+c_2x+c_3=0.
\end{equation}
Since $f(0)=c_3=-D^2<0$, and $\lim_{x\to+\infty}f(x)=+\infty$, we conclude that equation $h(z)$ always has at least one real positive root. Without loss of generality, suppose equation (\ref{Heq}) has three distinct positive roots, denoted y $x_1$, $x_2$ and $x_3$, respectively. Then, correspondingly, equation (\ref{OmEq}) also has three positive roots
\[
\omega_1=\sqrt{x_1},\hspace{0.3cm}\omega_2=\sqrt{x_2},\hspace{0.3cm}\omega_3=\sqrt{x_3}.
\]
From (\ref{ReIm})  one can find
\[
\cos\omega_k\tau=\frac{\omega^2(AD-CB+C\omega_k^2)}{D^2+C^2\omega^2},\hspace{0.3cm}k=1,2,3.
\]
If we denote
\begin{equation}
\displaystyle{\tau_k^{(n)}=\frac{1}{\omega_k}\left\{\arccos\left[\frac{\omega_k^2\left(AD-CB+C\omega_k^2\right)}{D^2+C^2\omega_k^2}\right]+2\pi n\right\},}\hspace{0.3cm}k=1,2,3,\hspace{0.3cm}n=0,1,2,...,
\end{equation}
then $\pm i\omega_k$ is a pair of purely imaginary roots of (\ref{CE}) with $\tau=\tau_k^{(n)}$. Since $\lim_{n\to\infty}\tau_k^{(n)}=\infty$, $k=1,2,3$, we can define
\begin{equation}\label{tau0}
\tau_0=\tau_{k_0}^{(0)}=\min_{k\in\{1,2,3\}}\{\tau_k^{(0)}\},\hspace{0.3cm}\omega_0=\omega_{k_0}.
\end{equation}
Using the results of \cite{RW2}, we can conclude that all roots of the characteristic equation (\ref{CE}) have negative real parts when $\tau\in[0,\tau_0)$. By construction, it follows that $f'(\omega_0^2)>0$, and this implies that at $\tau=\tau_0$, one has $\pm i\omega_0$ as a pair of simple purely imaginary roots of equation (\ref{CE}), see \cite{RW1} for details of the proof. The next step is that show that, in fact,
\[
\left[\frac{d{\rm Re}(\lambda(\tau))}{d\tau}\right]\Bigg|_{\tau=\tau_0}>0.
\]
To do this, we differentiate both sides of equation (\ref{CE}) with respect to $\tau$, which yields
\[
\frac{d\lambda(\tau)}{d\tau}=-\frac{\left(C\lambda^2+D\lambda\right)e^{-\lambda\tau}}{3\lambda^2+2A\lambda+B+(C-C\tau\lambda-\tau D)e^{-\lambda\tau}}.
\]
Evaluating the real part of this expression at $\tau=\tau_0$ and substituting $\lambda=i\omega_0$ gives
\begin{equation}
\left[\frac{d{\rm Re}(\lambda(\tau))}{d\tau}\right]\Bigg|_{\tau=\tau_0}=\frac{\omega_0^2\left[3\omega_0^4+2\omega_0^2(A^2-2B)+B^2-C^2\right]}{\Delta},
\end{equation}
where
\[
\Delta=\left(-3\omega_0^2+B-\tau_0 A\omega_0^2+C\cos\omega_0\tau_0\right)^2+\left(2A\omega_0+\tau_0\omega_0(B-\omega_0^2)-
C\sin\omega_0\tau_0\right)^2.
\]

\newpage
\begin{figure}[h]
\epsfig{file=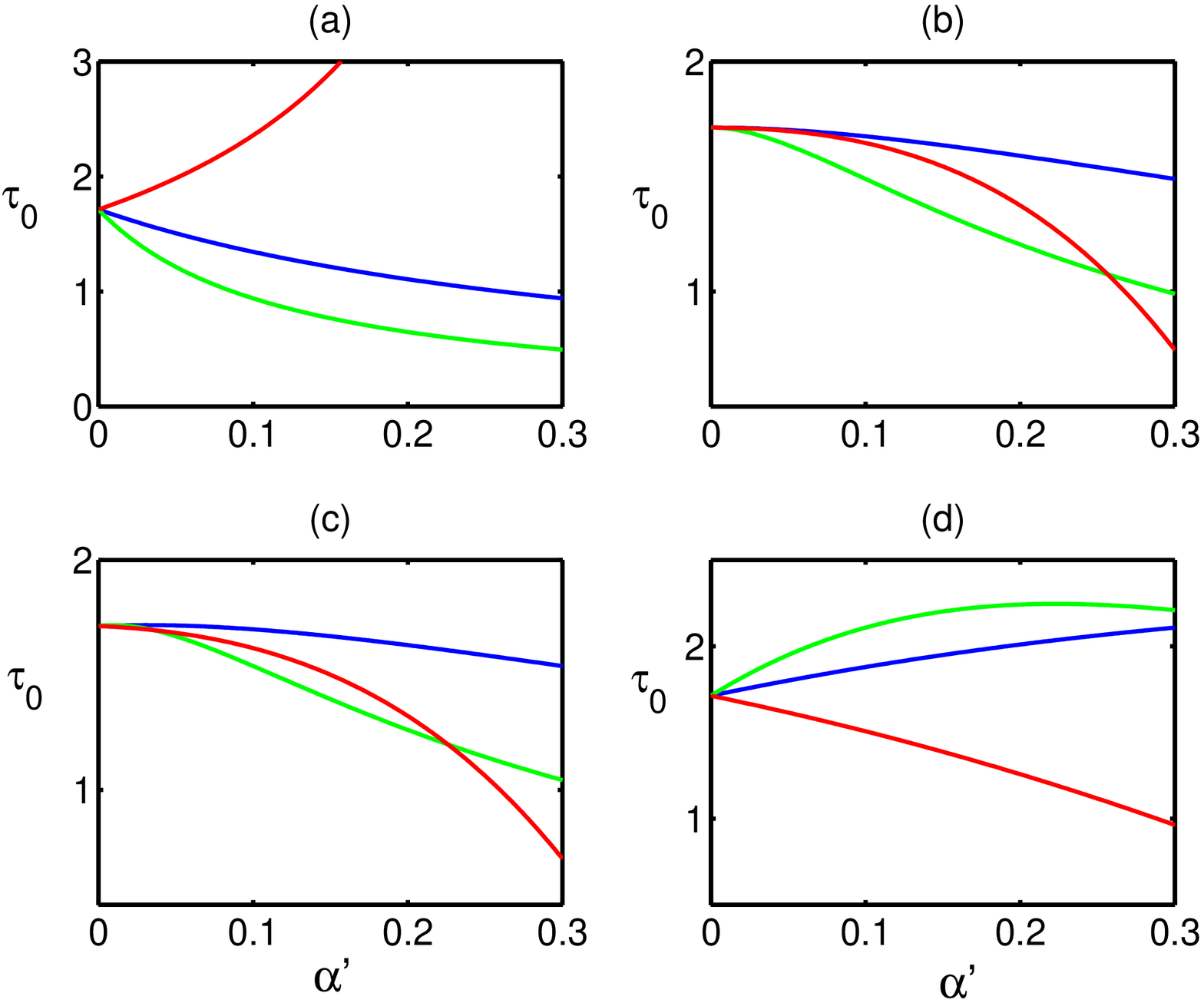,width=16cm}
\caption{Critical time delay $\tau_0$ (\ref{tau0}) at the Hopf bifurcation as a function of $\alpha'$. Parameter values are $\beta=0.2$, $\alpha=0.3$, $\mu=0.1$, $\beta'=0.8$, (a) $\mu'=0.1$, (b) $\mu'=0.24$, (c) $\mu'=0.255$, (f) $\mu'=0.7$. The colour corresponds to $B'=\beta'$ (blue), $B'=3\beta'$ (green), and $B'=-\beta'$ (red).}\label{LDfig}
\end{figure}
\noindent Using the definition of function $f$ from (\ref{fdef}) with the coefficients given in (\ref{Cpars}), we can alternatively write
\begin{equation}
\left[\frac{d{\rm Re}(\lambda(\tau))}{d\tau}\right]\Bigg|_{\tau=\tau_0}=\frac{\omega_0^2 f'(\omega_0^2)}{\Delta}.
\end{equation}
Since $f(0)<0$ and $\omega_0$ is the smallest positive root of (\ref{OmEq}), it follows that
\[
f'(\omega_0^2)>0,
\]
unless $\omega_0$ is a double root, in which case we take $\omega_0$ as the next root. Therefore, one concludes that
\[
\left[\frac{d{\rm Re}(\lambda(\tau))}{d\tau}\right]\Bigg|_{\tau=\tau_0}=\frac{\omega_0^2 f'(\omega_0^2)}{\Delta}>0,
\]
which, in the light of Hopf theorem, implies that at $\tau=\tau_0$, the fully symmetric steady state undergoes a Hopf bifurcation. We summarize these results as follows.\\

\noindent {\bf Theorem 2.} {\it The fully symmetric steady state $E$ is stable for $\tau<\tau_0$, where $\tau_0$ is defined in (\ref{tau0}), unstable for $\tau>\tau_0$, and undergoes a Hopf bifurcation at $\tau=\tau_0$. If the minimal value of $\tau_0$ corresponds to $B'=3\beta'$, the bifurcating solution will be an even periodic orbit; if it corresponds to $B'=-\beta'$, the bifurcating solution will be an odd periodic orbit; if it corresponds to $B'=\beta'$, the bifurcating periodic orbit will lie in the $V_4$ subspace.}\\

Figure~\ref{LDfig} illustrates how the time delay affects possible types of bifurcating solutions. Similar to the case of small time delay, the steady state $E$ bifurcates primarily into an even periodic orbit for smaller values of $\mu'$ and into an odd periodic orbit for higher values of $\mu'$. The main difference from the case of small time delay is that now the range of admissible values $\alpha'$ is bounded by $\alpha$ rather than the value of $\alpha'_H$ at the Hopf bifurcation for $\tau=0$. There are two main conclusions to be drawn from this Figure. The first one is that the computations of $\tau_0$ suggest that the state $E$ can only bifurcate to either even or odd periodic orbit, and the bifurcation into a subspace $V_4$ corresponding to two pairs of complex conjugate eigenvalues crossing the imaginary axis simultaneously does not happen, thus excluding the occurrence of vertex or edge oscillations, as well as discrete travelling waves \cite{Swift}. This, however, does not preclude them from appearing in the dynamics altogether, as they can arise as solutions bifurcating from the even/odd periodic orbits, as will be shown in the next section. The second conclusion which follows from Fig.~\ref{LDfig} is that as the efficiency of cross-reactive immune response $\alpha'$ increases, this leads to a decrease in $\tau_0$, thus leading to the onset of sustained oscillations for faster immune responses.

\section{Numerical simulations}

\begin{figure}[h]
\epsfig{file=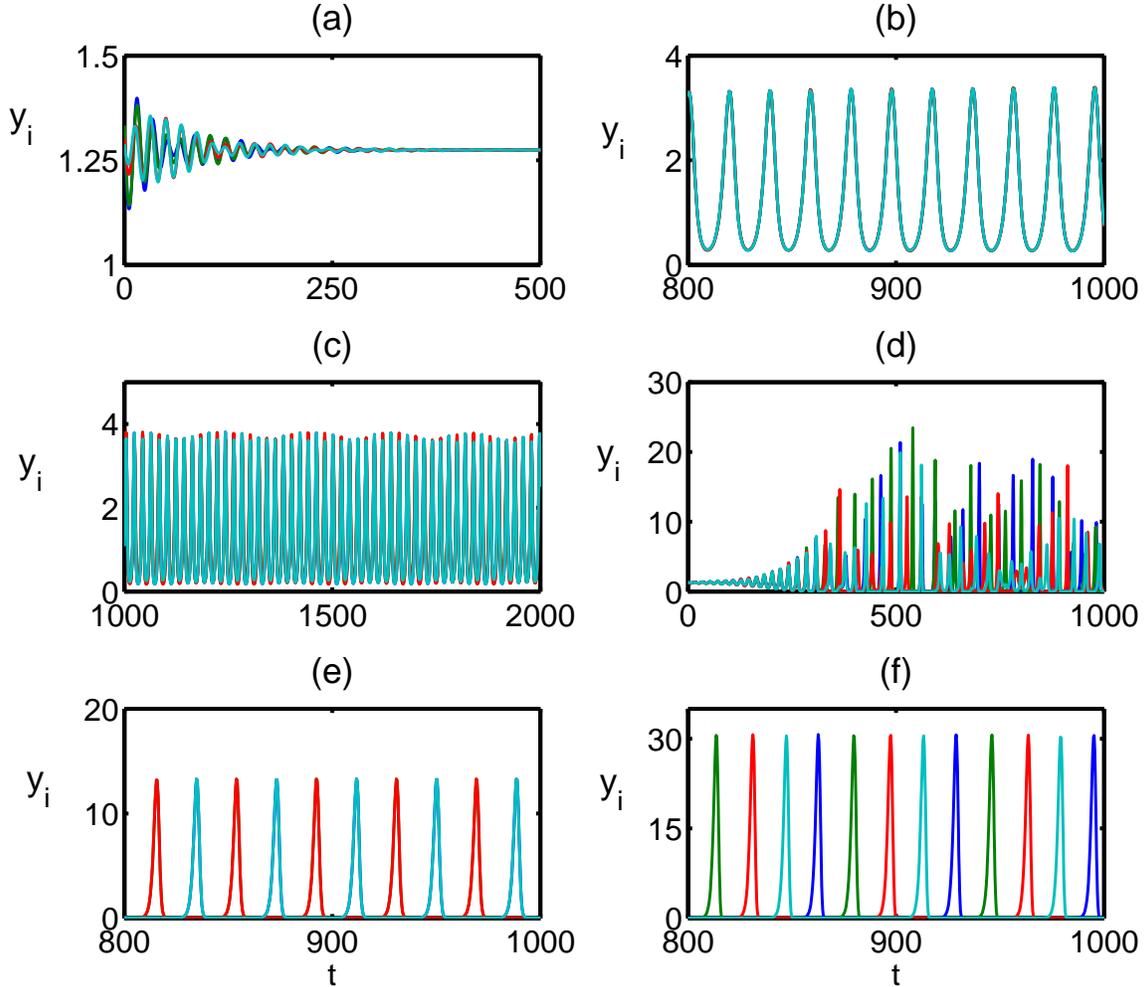,width=17cm}
\caption{Numerical solution of the system (\ref{vs}) with a connectivity matrix (\ref{Amat}). Parameter values are $\phi=1$, $\beta=0.2$, $\alpha=0.3$, $\mu=0.1$, $\beta'=0.8$. Different colours correspond to different antigenic variants. (a) Stable steady state $E$ ($\mu'=0.26$, $\alpha'=0.1$, $\tau=1$). (b) Even periodic orbit ($\mu'=0.26$, $\alpha'=0.1$, $\tau=1.26$). (c) Quasi-periodic even orbit ($\mu'=0.26$, $\alpha'=0.1$, $\tau=1.27$). (d) Chaotic solution ($\mu'=0.26$, $\alpha'=0.1$, $\tau=1.5$). (e) Odd periodic orbit ($\mu'=0.4$, $\alpha'=0.2$, $\tau=1.2$). (f) Discrete travelling wave ($\mu'=0.7$, $\alpha'=0.2$, $\tau=1.8$).}\label{TSols}
\end{figure}

In the previous sections we established that the fully symmetric steady state $E$ can undergo Hopf bifurcation, giving rise to two different types of solutions: a symmetric solution, in which all variants are oscillating identically, and an odd periodic orbit having variants 1 and 3 oscillating synchronously and half a period out-of-phase with variants 2 and 4. To understand evolution of these solutions as the time delay increases, we have performed a number of numerical simulations of system (\ref{vs}), which are shown in Fig.~\ref{TSols}. One can observe that for sufficiently small immune delay, the fully symmetric steady state $E$ is stable, see Fig. (a). As the time delay $\tau$ exceeds its critical value (\ref{tau0}) at the Hopf boundary, this state becomes unstable and gives rise to a stable fully symmetric periodic orbit, as demonstrated in Fig. (b). As the time delay increases, this solution acquires subharmonics, as shown in Fig. (c), and it eventually becomes chaotic, see Fig. (d). For other values of parameters, the fully symmetric steady state $E$ bifurcates into an odd periodic solution (see Fig. (e)), which for higher values of $\tau$ transforms into a discrete travelling wave, where each of the variants is quarter of period out-of-phase with its neighbours on the diagramme (\ref{Vard}). In fact, as the time delay is varied, it is possible to observe other stable periodic solutions with different phase shifts between antigenic variants.

The important difference from the model without time delay is that the only possibility for $\tau=0$ is a bifurcation into an odd periodic orbit \cite{B11}, while for $\tau>0$ there is also a possibility of the steady state $E$ bifurcating to a stable fully symmetric periodic orbit. In this case, Hopf bifurcation does not lead to the breakdown of the original ${\bf D}_4$ symmetry. Also, we note an important difference from the globally coupled system with delayed immune response. When considering a network of antigenic variants with an all-to-all coupling, Carr and Mitchell \cite{MC2} have shown that for the majority of parameter values, anti-phase Hopf bifurcation eventually leads to the behaviour that appears chaotic, while the simulations shown in Fig.~\ref{TSols} suggest that when not all antigenic variants are related to each other, the system is able to support a number of different out-of-phase solutions without going into chaotic regime.

To better understand the structure of different types of periodic behaviour in the model, one can use the so-called $H/K$ Theorem, which takes into account individual spatial and spatio-temporal
\newpage
\noindent symmetries of the solutions \cite{BuG,GSSym}. To use this method, we note that due to $\Gamma$-equivariance of the system (\ref{vs}) and uniqueness of its solutions, it follows that for any $T$-periodic solutions $x(t)$ and any element $\gamma\in\Gamma$ of the group, one can write
\[
\gamma x(t)=x(t-\theta),
\]
for some phase shift $\theta\in{\bf S}^{1}\equiv\mathbb{R}/\mathbb{Z}\equiv[0,T)$. The pair $(\gamma,\theta)$ is called a {\it spatio-temporal symmetry} of the solution $x(t)$, and the collection of all spatio-temporal symmetries of $x(t)$ forms a subgroup $\Delta\subset\Gamma\times{\bf S}^{1}$. One can identify $\Delta$ with a pair of subgroups, $H$ and $K$, such that $K\subset H\subset\Gamma$. We also define
\[
\begin{array}{l}
H\hspace{0.3cm}=\hspace{0.3cm}\left\{\gamma\in\Gamma:\gamma\{x(t)\}=\{x(t)\}\right\}\hspace{0.5cm}\mbox{spatio-temporal symmetries},\\
K\hspace{0.3cm}=\hspace{0.3cm}\left\{\gamma\in\Gamma:\gamma x(t)=x(t)\hspace{0.3cm}\forall t\right\}\hspace{0.6cm}\mbox{spatial symmetries.}
\end{array}
\]
Here, $K$ consists of the symmetries that fix $x(t)$ at each point in time, while $H$ consists of the symmetries that fix the entire trajectory. Under some generic assumptions on $H$ and $K$, the $H/K$ Theorem states that periodic states have spatio-temporal symmetry group pairs $(H,K)$ only if $H/K$ is cyclic, and $K$ is an isotropy subgroup \cite{BuG,GSSym}. The $H/K$ Theorem was originally derived in the context of equivariant dynamical systems by Buono and Golubitsky \cite{BuG}, and it has subsequently been used to classify various types of periodic behaviours in systems with symmetry that arise in a number of contexts, from speciation \cite{Spec} to animal gaits \cite{PG} and vestibular system of vertebrates \cite{Vest}.

In the case of  ${\bf D}_4$ symmetry group acting on four elements, there are eleven pairs of subgroups $H$ and $K$ satisfying the above requirements \cite{GSSym}. While periodic solutions corresponding to each such pair can exist in a general setup, the above theorem does not guarantee their existence or stability in a particular system, see discussion in \cite{GSSym}. Therefore, we use numerical simulations shown in in Fig.~\ref{TSols} to identify specific types of periodic solutions and their spatio-temporal symmetries that can be found in the system (\ref{vs}). Solutions shown in Figs. (b) and (c) have the full original symmetry of the system, and therefore are characterized by a pair $(H,K)=({\bf D}_4,{\bf D}_4)$. The solution shown in Fig. (d) is chaotic and does not have any of the symmetries of the original system. Figure (e) illustrates an odd periodic orbit with a symmetry $(H,K)=({\bf D}_4,{\bf D}_{2}^{p})$, where $K={\bf D}_{2}^{p}$ is an isotropy subgroup associated with reflections along the diagonals of the square. Finally, the solution shown in Fig. (f) is a discrete travelling wave with the symmetry $(H,K)=({\bf Z}_4,{\bf 1})$, also known as a "splay state" \cite{SM}, "periodic travelling (or rotating) wave" \cite{AKS}, or "ponies on a merry-go-round" or POMs \cite{AGM} in the studies of systems of coupled oscillators. In this dynamical regime all variants appear sequentially one after another along the diagramme in Fig.~(\ref{Vard}) with quarter of a period difference between two neighbouring variants. From the perspective of equvariant bifurcation theory, this solution is generic since the group ${\bf Z}_{n}$ is always one of the subgroups of the ${\bf D}_n$ group for the ring coupling,  or the ${\bf S}_n$ group for an all-to-all coupling, and its existence has already been extensively studied \cite{AGM,GS2,GSS}. From the immunological point of view, this is an extremely important observation that effectively such solution, which immunologically represents sequential appearance of parasitemia peaks corresponding to different antigenic variants, owes its existence not to the individual dynamics of antigenic variants, but rather to the particular symmetric nature of cross-reactive interactions between them. This immunological genericity ensures that the same conclusions hold for a wide variety of immune interactions between human host and parasites, which use antigenic variation as a mechanism of immune escape, as illustrated, for instance, by malaria, African Trypanosomes, several members of {\it Neisseria} family ({\it N. meningitidis} and {\it N. gonorrhoeae}), {\it Borrelia hermsii} etc. \cite{Gu,Tur}.

\section{Discussion}

In this paper we have used methods of equivariant bifurcation theory to understand the effects of immune delay on the dynamics in a model of antigenic variation in malaria. In the simplest case of two epitopes with two variants in each epitope, the system is equivariant with respect to a ${\bf D}_4$ symmetry group of the square. Using isotypic decomposition of the phase space based on the irreducible representations of this symmetry group has allowed us to find critical value of the time delay at the boundary of Hopf bifurcation of the fully symmetric steady state in terms of system parameters. We have identified even and odd periodic solutions that can arise at Hopf bifurcation, and also performed numerical simulations of the full system to illustrate other types of dynamical behaviours that can be exhibited by the model. These have been classified in terms of their spatial and spatio-temporal symmetries using the $H/K$ Theorem. Our analysis suggests that as the efficiency of the cross-reactive immune response increases, the critical value of the time delay at the Hopf bifurcation decreases, which is similar to earlier studied cases of complete synchrony \cite{RG,MC1} or global coupling between antigenic variants \cite{MC2}. At the same time, unlike these , out system is able to support a range of stable phase shift solutions without developing chaotic dynamics.

When applying the results of our analysis to the studies of realistic models of antigenic variation, one of the important consideration that have to be taken into account is the fact that in reality systems of antigenic variants do not always fully preserve the symmetry assumed in the mathematical models. In the context of modelling immune interactions between distinct antigenic variants, this means that not all variants cross-react with each other in exactly the same quantitative manner. Despite this limitation, due to the normal hyperbolicity, which is a generic property in such models, main phenomena associated with the symmetric model survive under perturbations, including symmetry-breaking perturbations. The discussion of this issue in the context of modelling sympatric speciation using a symmetric model can be found in \cite{GSSym}.

The results presented in this paper are quite generic, and the conclusions we obtained are valid for a wide range of mathematical models of antigenic variation. In fact, they are applicable to the analysis of within-host dynamics of any parasite, which exhibits similar qualitative features of immune interactions based on the degree of relatedness between its antigenic variants. The significance of this lies in the possibility to classify expected dynamical regimes of behaviour using very generic assumption regarding immune interactions, and they will still hold true, provided the actual system preserves the underlying symmetries. 

There are several ways in which the analysis in this paper can be further improved to achieve an even more realistic representation of the dynamics of antigenic variation. One of the assumptions in our analysis is that that the degree of cross-reactivity between antigenic variants does not vary with the number of epitopes they share. It is straightforward, however, to introduce antigenic distance between antigenic variants in a manner similar to the Hamming distance \cite{AdSa,RG05}. Such a modification would not alter the topology of the network of immune interactions, but rather it would assign different weights to connections between different antigenic variants in such a network. Another modelling issue concerns the way how the time delay in the immune response can be represented mathematically in the most realistic manner. It would be beneficial to investigate the dynamics of antigenic variation under the influence of a more general distributed time delay, which is known to cause both destabilization of steady states, and also suppression of oscillations \cite{AvD,BK,KBS,L}. This would provide better insights into how the efficiency of developing and maintaining immune response affects the within-host dynamics of parasites with antigenic variation.  Alternatively, one can introduce different time delays for the development of specific and long-lasting immune responses and analyse how the difference in these timescales affects the overall stability and dynamics. Systematic analysis of the effects of the introduction of antigenic distances and different/distributed delays in the immune response on the dynamics of antigenic variation is the subject of further study.

\end{document}